\documentclass[12pt]{article}

\usepackage{amsmath}
\usepackage{amssymb}
\usepackage{graphicx}
\usepackage{url}
\usepackage{multirow}
\usepackage{authblk}
\usepackage{mathtools}
\usepackage{hyperref}
\usepackage[margin=1in]{geometry}
\usepackage{natbib}
\usepackage{subfig}
\usepackage{graphicx}

\usepackage{adjustbox}

\usepackage{float}

\newcommand\independent{\protect\mathpalette{\protect\independenT}{\perp}}
\def\independenT#1#2{\mathrel{\rlap{$#1#2$}\mkern2mu{#1#2}}}

\usepackage{tikz}

\usetikzlibrary{shapes.geometric, arrows}
\tikzstyle{box} = [minimum width=1.0cm, minimum height=1.0cm, text width=2.5cm, text centered, draw=black,fill=green!10]
\tikzstyle{oval} = [minimum width=1.0cm, minimum height=1.0cm, text width=1.75cm, text centered, draw=black,fill=red!10]
\tikzstyle{arrow} = [thick,->,>=stealth]

\begin{document}

\title{Causal Mediation Analysis Decomposition of Between-hospital Variance}

\author[1]{ Bo~Chen}
\author[2]{Keith A.~Lawson}
\author[2]{Antonio~Finelli}
\author[1]{Olli~Saarela\thanks{Correspondence to: Olli Saarela, Dalla Lana School of Public Health, 155 College Street, Toronto, Ontario M5T 3M7, Canada. Email: \texttt{olli.saarela@utoronto.ca}}}

\affil[1]{Dalla Lana School of Public Health, University of Toronto}
\affil[2]{Princess Margaret Cancer Centre, University Health Network}

\maketitle

\begin{abstract}
Causal variance decompositions for a given disease-specific quality indicator can be used to quantify differences in performance between hospitals or health care providers. While variance decompositions can demonstrate variation in quality of care, causal mediation analysis can be used to study care pathways leading to the differences in performance between the institutions. This raises the question of whether the two approaches can be combined to decompose between-hospital variation in an outcome type indicator to that mediated through a given process (indirect effect) and remaining variation due to all other pathways (direct effect). For this purpose, we derive a causal mediation analysis decomposition of between-hospital variance, discuss its interpretation, and propose an estimation approach based on generalized linear mixed models for the outcome and the mediator. We study the performance of the estimators in a simulation study and demonstrate its use in administrative data on kidney cancer care in Ontario.

\end{abstract}

\newpage

\section{Introduction}\label{section:intro}

Quality of healthcare can be compared between institutions such as hospitals or individual health care providers using disease specific quality indicators (QIs). These measure structural, process or outcome elements related to the care of a particular condition \citep{donabedian:1988}. The resulting comparison in terms of outcome type measures is commonly referred to as hospital/provider profiling; with statistical methods for this discussed for example by \citet{goldstein:1996,racz:2010}. The comparisons typically require case-mix or risk adjustment to account for different patient populations treated by the different institutions/providers, with common methods reviewed by \citet{Shahian:2008}. Such comparisons can also be framed counterfactually in a causal inference framework, comparing the quality of care the patient would receive if treated by a different institution/provider. For instance, \citet{varewyck:2014} used the potential outcomes framework \citep{rosenbaum:1983} to represent patient outcomes indexed by hospital, considering the hospital index as a multicategory categorical treatment/exposure variable. The resulting standardized hospital-specific mean outcomes can then be compared pairwise, ranked, or compared to average performance in the health care system. To reduce the number of comparisons, \citet{Chen:2019} proposed a causal variance decomposition which characterizes the overall between-hospital variation through a variance component with a specific causal interpretation. Variance decompositions in profiling health care providers have also been considered outside the causal modeling framework by e.g. \citet{xia2020accounting}. While variance decompositions can demonstrate differences between the hospitals, causal mediation analysis can be used to study care pathways leading to the differences in performance between the hospitals. In particular, if between-hospital variation is found in an outcome type indicator, this raises the question of whether this can be explained by between-hospital variation in a process type indicator.  For instance, \citet{Daignault:2019} explained the observed variation of length of the stay after radical nephrectomy for early-stage (T1-T2) kidney cancer patients by between-hospital variation in minimally invasive surgery (MIS) rates, with the pathway through MIS understood as the indirect (mediated) effect, and all other pathways as the direct effect.

The potential outcomes mediation analysis framework \citep[e.g.][]{VanderWeele:2009b, VanderWeele:2014} can be adapted to between-hospital comparisons by decomposing each pairwise comparison between two hospitals into natural indirect effect and natural direct effect. However, with a large number of hospitals, the pairwise comparisons may not all be of interest. Alternatively, it would be possible to decompose the contrast to the average performance, reducing the number of comparisons. Indirect standardization is commonly used in hospital comparisons, with the standardized mortality/morbidity ratio (SMR) used as the effect measure. \citet{Daignault:2019} proposed a causal mediation analysis decomposition for the SMR. The difference to a directly standardized quantity is that it uses the entire combined patient population as the standard population, whereas with SMR each comparison is conditional on the patient population of the given hospital. The number of the comparisons in the SMR mediation analysis is equal to the number of the number of hospitals, and while this is less than the number of pairwise hospital comparisons, there may still be limited statistical power to detect mediation, especially for small volume hospitals. This motivates the question of whether between-hospital variance in an outcome type indicator can be decomposed in causal mediation analysis sense. 

In the psychometric literature, mediation analysis is usually based on the classical \citet{Reuben:1986} approach. In that field, several different effect size and variance explained/$R^2$ type measures have been proposed for quantifying mediation in the linear structural equation modeling framework \citep{fairchild2009r,de2012r,miovcevic2018statistical,lachowicz2018novel}. The general challenge in quantifying mediation in terms of variance explained is that both direct and indirect effect are due to the treatment/exposure variable, making them dependent since they have to add up to the total effect \citep{de2012r}. In principle, the total effect variance explained can be decomposed, but the decomposition has three terms, including a covariance type term between the direct effects and the indirect effects. This has motivated a number of single number effect measures that are not directly based on such a decomposition, reviewed by \citet{miovcevic2018statistical} and \citet{lachowicz2018novel}. However, because these proposals are specific to the linear structural equation framework, we proceed to further consider the three-way decomposition of the total effect variance explained. We will argue that this is meaningful especially with multi-category categorical exposures. However, to understand the causal interpretation of the decomposition and to generalize it, we have to derive it in the potential outcomes framework. Because we aim for decomposition on the scale of the outcome variable, our approach will also allow incorporation of link functions and exposure-mediator interactions.

We make a distinction between the present framework, and methods that have been proposed for multilevel mediation analysis \citep{preacher2010general,tofighi2014single,zigler2019comparison}, which refers to mediation analysis in the presence of clustered data, in particular when some of the variables are measured at the cluster rather than at individual level. This is different from the present problem, where although clusters are present as the hospitals, they are directly used as the categorical exposure variable. In our case the mediator and outcome variable, which are patient outcomes and processes of care used to construct QIs, are measured at individual level, and thus we aim for individual-level causal interpretation. We would be in the multilevel mediation setting if we were interested in some structural characteristic of the hospitals, such as academic affiliation, as the exposure, but we do not consider this here. However, in the discussion we consider briefly how our proposal can be generalized in the presence of multiple layers of exposures, such as surgeons within hospitals.

Based on the objectives motivated above, the structure of the paper is as follows. In Section \ref{review}, we review the previous measures in the mediation analysis. In Section \ref{Notation and assumptions}, we introduce the potential outcomes (Rubin's) causal model and its extension to mediation analysis and relevant assumptions. In Section \ref{Causal mediation decomposition}, we review our previously proposed three-way causal decomposition for the observed variation in care received, and propose the causal mediation analysis decomposition of the between-hospital variance. We further investigate its interpretation and special cases in Sections \ref{interpretation} and \ref{Hypothetical}. We propose model-based estimation methods in Section \ref{section:estimation}, and investigate their performance through a simulation study and illustrate their use in a real data analysis in Sections \ref{section:simulation} and \ref{section:illustration}. Finally, we discuss the limitations and future research directions in Section \ref{section:discussion}.

\section{Previously proposed measures}\label{review}

\subsection{Natural direct and indirect effects}

We first introduce the necessary notation. Let $Y$ be the observed real-valued outcome used to construct a QI (say, log length of stay after radical nephrectomy), $Z \in \{1, \ldots, q\}$ indicator for the hospital where the patient received treatment, $M\in \{0,1\}$ and the observed binary mediator (say, indicator for minimally invasive surgery versus open surgery), and $X=(X_{1}, \ldots, X_{p})$ be a vector of patient case-mix variables. The potential version of the mediator is $M_z$, giving the mediator level had the patient been treated in hospital $z \in \{1, \ldots, q\}$. Similarly, $Y_{zm}$ is the potential outcome of the same patient received surgical treatment $m \in \{1,0\}$ via hospital $z$. $Y_{zM_{z^*}}$ is the potential outcome had the patient been treated in $z$ but setting the mediator to the level it would have taken in hospital $z^*$, without fixing it to a specific value. The total effect of the comparison of an index hospital $z$ to a reference hospital $z^*$ can be decomposed as
\begin{align*}
TE_{z,z^*}&=E[Y_{zM_{z}}]-E[Y_{z^*M_{z^*}}]\\
&=(E[Y_{zM_{z}}]-E[Y_{zM_{z^*}}]) + (E[Y_{zM_{z^*}}]-E[Y_{z^*M_{z^*}}])\\
&=NIE_{z,z^*}+NDE_{z,z^*},
\end{align*}
where TE, NIE, and NDE represent the total effect, natural indirect effect, and natural direct effect, respectively. The expected potential outcomes can be estimated through the mediation formula, resembling direct standardization. Due to the multi-category categorical exposure, there are $q(q-1)/2$ pairwise comparisons (e.g. 3160 pairwise comparisons when $q=80$). Alternatively, it would be possible to decompose the contrast to the average performance $E[Y_{zM_{z}}]-E[Y_{ZM_{Z}}] = E[Y_{zM_{z}}]-E[Y]$, as
\begin{align*}
TE_{z}&=E[Y_{zM_{z}}]-E[Y_{ZM_{Z}}]\\
&=(E[Y_{zM_{z}}]-E[Y_{zM_{Z}}]) + (E[Y_{zM_{Z}}]-E[Y_{ZM_{Z}}])\\
&=NIE_{z}+NDE_{z},
\end{align*}
resulting in $q$ comparisons. An alternative decomposition with a slightly different interpretation could be obtained by introducing the term $E[Y_{ZM_{z}}]$ instead of $E[Y_{zM_{Z}}]$ in the second equality.  \citet{Daignault:2019} proposed a causal mediation analysis decomposition for the SMR as follows. Let $A\in \{1, \ldots, q\}$ be a hypothetical ``randomized'' target assignment regime used for random draws of potential outcomes. The SMR can be decomposed as
\begin{align*}
SMR_z^{TE}&=\frac{E[Y_{zM_z}\mid Z=z]}{E[Y_{AM_A}\mid Z=z]}\\
&=\frac{E[Y_{zM_z}\mid Z=z]}{E[Y_{zM_A}\mid Z=z]} \times\frac{E[Y_{zM_A}\mid Z=z]}{E[Y_{AM_A}\mid Z=z]}\\
&=SMR_z^{NIE}\times SMR_z^{NDE},
\end{align*}
where $SMR_z^{TE}, SMR_z^{NIE}$, and $SMR_z^{NDE}$ represent the total effect, the natural indirect effect, and the natural direct effect SMR, respectively.

\subsection{Mediation analysis decomposition of $R^2$}

In the linear modeling framework, \citet{de2012r} considered mediation analysis decomposition of $R^2$. Taking $Z$ to be dichotomous, specify the three linear regression models
\begin{align*}
E[Y \mid Z=z, X=x]=\alpha_1+\beta_{1}z+\gamma_1x,
\end{align*}
\begin{align}\label{3linear}
E[Y \mid Z=z, M=m, X=x]=\alpha_2+\beta_{2}z+\beta_{3}m+\gamma_2x,
\end{align}
\begin{align*}
E[M \mid Z=z, X=x]=\alpha_3+\beta_{4}z+\gamma_3x.
\end{align*}
In the \citet{Reuben:1986} approach, assuming that conditioning on $X$ is sufficient to control for confounding, the regression coefficient $\beta_1$ represents the total effect of $Z$ on $Y$, $\beta_2$ represents the direct effect of $Z$ on $Y$, and $\beta_3\beta_4$ or $\beta_1-\beta_2$ represents the indirect effect of $Z$ on $Y$ via $M$ (product and subtraction methods). 

If the variables have been standardized, the total variance of $Y$ explained by $Z$ is the square of the regression coefficient $\beta_1 = \beta_{total}$. For this, one can write
\begin{align}\label{R2_type}
R^2 = \beta_{total}^2=(\beta_{direct}+\beta_{indirect})^2=\beta_{direct}^2+\beta_{indirect}^2+2\beta_{direct}\beta_{indirect}.
\end{align}
Here the terms $\beta_{direct}^2=\beta_2^2$ and $\beta_{indirect}^2=(\beta_3\beta_4)^2$ are due to the direct and indirect effects of $Z$, but the presence of the third term 
$\beta_{direct}\beta_{indirect}$, which can also be negative, means that the dependency of the direct and indirect effects cannot be ignored \citep{de2012r}.

\section{Proposed measures}

\subsection{Motivating context}\label{Context}

\begin{figure}[!ht]
\centering
\includegraphics[width=0.8\textwidth]{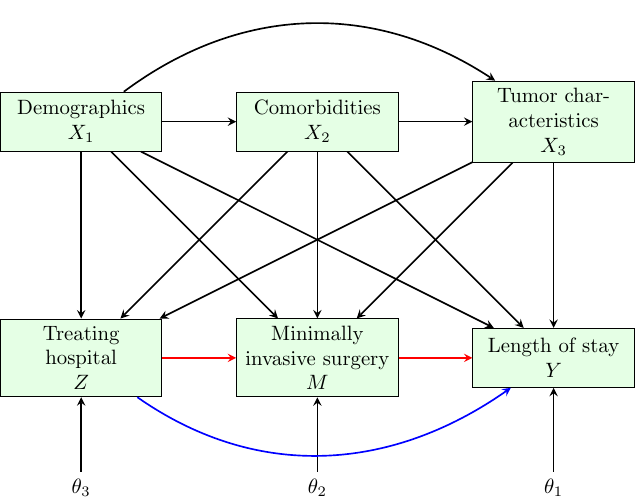}
\caption{Example DAG where a process of care is mediating the hospital effect on outcome. The arrows $Z \rightarrow M$ and $M \rightarrow Y$ reflect the part of the hospital effect that is mediated by the process (indirect effect), while the arrow $Z \rightarrow Y$ reflects the hospital effect due to all other causal pathways (direct effect).}
\label{figure:dag}
\end{figure}

We will discuss the concepts in the context of the running example where $Y$ is (log-transformed) length of stay after radical nephrectomy used as primary treatment for early stage kidney cancer, $M$ is minimally invasive (laparoscopic or robotic) vs open surgical approach, and $Z$ is the hospital index. However, generally for the subsequent development, the outcome and mediator can be any of dichotomous, count or continuous variables. We will revisit some of these situations in Section \ref{section:estimation}. The hypothesized causal relationships are illustrated in the directed acyclic graph (DAG) in Figure \ref{figure:dag}. Here we are interested whether between-hospital variation in length of stay after radical nephrectomy for kidney cancer patients is explained by between-hospital variation in minimally invasive vs open surgery. Hence, we are interested in decomposing between-hospital variation in the length of stay to that mediated through surgical approach (indirect effect: $Z\rightarrow M \rightarrow Y$) and remaining variation due to all other pathways (direct effect: $Z\rightarrow Y$). The relevant covariates $X = (X_1,X_2,X_3)$ can be classified into the three categories shows in the DAG, namely demographic factors (e.g. age, sex), comorbidities (e.g. Charlson score) and tumor characteristics/disease progression (e.g. T-stage). These can all be already imbalanced between the treating hospitals (e.g. more complicated cases may be referred to high volume cancer centers), so they are considered potential confounders of all the considered relationships ($Z\rightarrow M$, $Z\rightarrow Y$ and $M\rightarrow Y$).

\subsection{Causal assumptions}\label{Notation and assumptions}

The observed variables $Y$ and $M$ are linked to their potential counterparts under the counterfactual consistency/stable unit treatment value assumption (SUTVA), by $Y= Y_Z=Y_{ZM_Z}$ and $M = M_Z$. Causal inferences on the controlled direct hospital effects on the outcome are possible under the assumption of strong ignorability of the joint hospital and mediator assignment mechanism, which states that $0 < P(Z = z, M = m \mid X = x) < 1$ for all $z \in \{1, \ldots, q\}$, $m \in \{1, 0\}$ and $x$ (positivity) and $Y_{zm} \independent  (Z,M) \mid X$ (conditional exchangeability). Additionally, we need to assume $M_z \independent  Z \mid X$ (conditional exchangeability with respect to the mediator) and $Y_{zm} \independent  M_{z^*} \mid X$ (no exposure-induced confounders of the mediator-outcome relationship) to identify the natural indirect and natural direct effects, as defined in Section \ref{review}. 

\subsection{Causal mediation analysis decomposition of between-hospital variance}\label{Causal mediation decomposition}

We first decompose the observed variance in outcome, $V[Y]$, to that causally explained by the differences between hospitals, by individual-level case-mix factors, and residual variation. Under the counterfactual consistency, $V[Y] = V[Y_Z] = V[Y_{ZM_Z}]$. We introduce the shorthand notations $e(z; X) = P(Z = z \mid X)$ for the hospital assignment probabilities. In \citet{Chen:2019}, we derived a three-way causal variance decomposition in care received,
\begin{align}\label{3-way}
V[Y] &= V_{X}\left\{\sum_{z}E(Y_z \mid X)e(z; X)\right\} \nonumber \\
&\quad + E_{X} \left \{ \sum_{z}\left[E(Y_z \mid X)  - \sum_{z^*} E(Y_{z^*} \mid X) e(z^*; X) \right]^2 e(z; X) \right \} \nonumber \\
&\quad + E_{X} \left \{ \sum_{z} V(Y_z \mid X) e(z; X) \right \}.
\end{align}
The first and third variance terms here represent the case-mix and residual variance, respectively. The second term represents the average squared difference from the average outcome in the health care system; we take this to be the causal quantity of interest representing between-hospital variation for similar patients. The inside of this term can be further expanded to introduce the mediator $M$ as
\begin{align}\label{mediation}
\MoveEqLeft  \sum_{z}\bigg[E(Y_z \mid X)  - \sum_{z^*} E(Y_{z^*} \mid X) e(z^*; X) \bigg]^2 e(z; X) \nonumber\\
&= \sum_{z}\bigg[E(Y_{zM_z} \mid X)  - \sum_{z^*} E(Y_{z^*M_{z^*}} \mid X) e(z^*; X) \bigg]^2 e(z; X) \nonumber\\
&= \sum_{z}\bigg\{ \bigg[E(Y_{zM_z} \mid X)  - \sum_{z^*} E(Y_{zM_{z^*}} \mid X) e(z^*; X) \bigg] \nonumber\\
&\quad+ \bigg[ \sum_{z^*} E(Y_{zM_{z^*}} \mid X) e(z^*; X) - \sum_{z^*} E(Y_{z^*M_{z^*}} \mid X)e(z^*; X) \bigg] \bigg\}^2 e(z; X) \nonumber\\
&= \sum_{z}\bigg[E(Y_{zM_z} \mid X)  - \sum_{z^*} E(Y_{zM_{z^*}} \mid X) e(z^*; X) \bigg]^2 e(z; X) \nonumber\\
&\quad+ \sum_{z} \bigg[ \sum_{z^*} E(Y_{zM_{z^*}} \mid X) e(z^*; X) - \sum_{z^*} E(Y_{z^*M_{z^*}} \mid X)e(z^*; X) \bigg]^2 \nonumber\\
&\quad\quad  e(z; X) \nonumber\\
&\quad+ \sum_{z}2 \bigg[E(Y_{zM_z} \mid X)  - \sum_{z^*} E(Y_{zM_{z^*}} \mid X) e(z^*; X) \bigg] \nonumber\\
&\quad\quad\times \bigg[ \sum_{z^*} E(Y_{zM_{z^*}} \mid X) e(z^*; X) - \sum_{z^*} E(Y_{z^*M_{z^*}} \mid X) e(z^*; X) \bigg] e(z; X).
\end{align}
Thus, the between-hospital variance can be decomposed into three parts, weighted average of the squared natural indirect effects (compared to the average level of the mediator across the hospitals), weighted average of the squared natural direct effects (hospital effects at the average level of the mediator), and a covariance type term between the indirect and direct effects. The last term can also be negative because the indirect and direct effects can work to opposite directions, making the hospitals more similar than the squared terms would imply. The last term is of interest in its own right because it can tell whether hospitals that perform poorly in terms of the mediator are also performing worse in other respects. 

\subsection{Causal interpretation of the decomposition}\label{interpretation}

We note that in the absence of (additive) hospital-mediator interaction the hospital $z$ specific indirect effect can be written as 
\begin{align*}
\MoveEqLeft E(Y_{zM_z} \mid X)  - \sum_{z^*} E(Y_{zM_{z^*}} \mid X) e(z^*; X) \\
&= \sum_{z^*} \left[E(Y_{zM_z} \mid X) -  E(Y_{zM_{z^*}} \mid X)\right] e(z^*; X) \\
&= \sum_{z^*} \left[E(Y_{z'M_z} \mid X) -  E(Y_{z'M_{z^*}} \mid X)\right] e(z^*; X) 
\end{align*}
and the direct effect as
\begin{align*} 
\MoveEqLeft \sum_{z^*} E(Y_{zM_{z^*}} \mid X) e(z^*; X) - \sum_{z^*} E(Y_{z^*M_{z^*}} \mid X)e(z^*; X) \\
&= \sum_{z^*} \left[E(Y_{zM_{z^*}} \mid X) - E(Y_{z^*M_{z^*}} \mid X)\right]e(z^*; X) \\
&= \sum_{z^*} \left[E(Y_{zM_{z'}} \mid X) - E(Y_{z^*M_{z'}} \mid X)\right]e(z^*; X) 
\end{align*}
for any $z' \in \{1, \ldots, q\}$. Thus, the expected (over $z$) indirect and direct effects become
\begin{align*}
\MoveEqLeft \sum_{z} \left\{ E(Y_{zM_z} \mid X)  - \sum_{z^*} E(Y_{zM_{z^*}} \mid X) e(z^*; X) \right\} e(z; X) \\
&= \sum_{z} \left\{ \sum_{z^*} \left[E(Y_{z'M_z} \mid X) -  E(Y_{z'M_{z^*}} \mid X)\right] e(z^*; X) \right\} e(z; X) \\
&= \sum_{z} E(Y_{z'M_z} \mid X) e(z; X) -  \sum_{z^*} E(Y_{z'M_{z^*}} \mid X) e(z^*; X) = 0
\end{align*}
and
\begin{align*}
\MoveEqLeft \sum_{z} \left\{ \sum_{z^*} E(Y_{zM_{z^*}} \mid X) e(z^*; X) - \sum_{z^*} E(Y_{z^*M_{z^*}} \mid X)e(z^*; X) \right\} e(z; X) \\
&=\sum_{z} \left\{ \sum_{z^*} \left[E(Y_{zM_{z'}} \mid X) - E(Y_{z^*M_{z'}} \mid X)\right]e(z^*; X) \right\} e(z; X) \\
&=\sum_{z}  E(Y_{zM_{z'}} \mid X) e(z; X) - \sum_{z^*} E(Y_{z^*M_{z'}} \mid X) e(z^*; X) = 0
\end{align*}
Thus, under the no-interaction assumption, the terms in the decomposition (\ref{mediation}) are directly the variance of the hospital-specific indirect effects, variance of the direct effects and two times the covariance of the indirect and direct effects. However, decomposition (\ref{mediation}) can be defined and estimated without making this assumption. In the presence of hospital-mediator interaction, because the hospital-specific direct and indirect effects are weighted averages of the pairwise indirect and direct effects, the interactions can contribute to both of these.

To understand the causal interpretation of the new decomposition, it is also helpful to consider the special case of two hospitals. With two hospitals indexed by $z=0,1$ and letting the propensity score $e(1; X)=\pi(X)$, the first term of the decomposition (weighted average of squared indirect effects) becomes
\begin{align*}
\MoveEqLeft \sum_{z=0}^1\bigg[E(Y_{zM_z} \mid X)  - \sum_{z^*=0}^1 E(Y_{zM_{z^*}} \mid X) e(z^*; X) \bigg]^2 e(z; X)\\
&=\pi(X)(1-\pi(X))\bigg[ [E(Y_{1M_1}\mid X)- E(Y_{1M_0}\mid X)]^2 (1-\pi(X))\\
&\quad+  [E(Y_{0M_1}\mid X)- E(Y_{0M_0}\mid X)]^2 \pi(X)\bigg]
\end{align*}
the second term of the decomposition (weighted average of squared direct effects) becomes
\begin{align*}
\MoveEqLeft  \sum_{z=0}^1 \bigg[ \sum_{z^*=0}^1 E(Y_{zM_{z^*}} \mid X) e(z^*; X) - \sum_{z^*=0}^1 E(Y_{z^*M_{z^*}} \mid X) e(z^*; X) \bigg]^2 e(z; X)\\
&=\pi(X)(1-\pi(X))\bigg[ [E(Y_{1M_0}\mid X)- E(Y_{0M_0}\mid X)]^2 (1-\pi(X))\\
&\quad+  [E(Y_{1M_1}\mid X)- E(Y_{0M_1}\mid X)]^2 \pi(X)\bigg]
\end{align*}
the last term of the decomposition (representing the covariance between the indirect and direct effects) becomes
\begin{align*}
\MoveEqLeft  \sum_{z=0}^1 2 \bigg[E(Y_{zM_z} \mid X)  - \sum_{z^*=0}^1 E(Y_{zM_{z^*}} \mid X) e(z^*; X) \bigg] \\
\MoveEqLeft \times \bigg[ \sum_{z^*=0}^1 E(Y_{zM_{z^*}} \mid X) e(z^*; X) - \sum_{z^*=0}^1 E(Y_{z^*M_{z^*}} \mid X) e(z^*; X) \bigg] e(z; X)\\
&=2\pi(X)(1-\pi(X))\\
&\quad\bigg[ [E(Y_{1M_1}\mid X)- E(Y_{1M_0}\mid X)] [E(Y_{1M_0}\mid X)- E(Y_{0M_0}\mid X)](1- \pi(X))\\
&\quad+ [E(Y_{0M_0}\mid X)- E(Y_{0M_1}\mid X)] [E(Y_{0M_1}\mid X)- E(Y_{1M_1}\mid X)]\pi(X)\bigg].
\end{align*}
In the special case of two hospitals, all of the three terms are functions of pairwise causal contrasts. For further interpretation, we consider three relevant scenarios based on the relationships between $Y , Z,  X$ and $M$ in Figure \ref{figure:dag}.
\begin{description}
\item[Scenario 1.] In the absence of the arrows $Z\rightarrow M$, which implies $Z\independent M\mid X$, we have that $E(Y_{1M_1}\mid X)=E(Y_{1M_0}\mid X)$ and $E(Y_{0M_1}\mid X)=E(Y_{0M_0}\mid X)$. The first and last terms are equal to 0, which implies that there is no indirect between-hospital effect on the length of stay via mediator. 
The between-hospital variance reduces to the second term
\begin{align*}
\pi(X)(1-\pi(X))\bigg[ &[E(Y_{1M_0}\mid X)- E(Y_{0M_0}\mid X)]^2 (1-\pi(X))\\
&+  [E(Y_{1M_1}\mid X) - E(Y_{0M_1}\mid X)]^2 \pi(X)\bigg],
\end{align*}
 which demonstrates that it is fully due to the direct between-hospital effect on the length of stay.
\item[Scenario 2.]  In the absence of the arrow $M\rightarrow Y$, which implies $Y\independent M\mid (Z, X)$, we have that $E(Y_{1M_1}\mid X)=E(Y_{1M_0}\mid X)$ and $E(Y_{0M_1}\mid X)=E(Y_{0M_0}\mid X)$. The interpretations of the three terms in the decomposition are the same as in Scenario 1.
\item[Scenario 3.] In the absence of the arrow $Z\rightarrow Y$, which implies $Y\independent Z\mid (M, X)$, we have that $E(Y_{1M_1}\mid X)=E(Y_{0M_1}\mid X)$ and $E(Y_{1M_0}\mid X)=E(Y_{0M_0}\mid X)$.  The second and last terms are equal to 0, which implies there is no direct between-hospital effect on the length of stay. 
The between-hospital variance reduces to the first term
\begin{align*}
\pi(X)(1-\pi(X))\bigg[ &[E(Y_{1M_1}\mid X)- E(Y_{1M_0}\mid X)]^2 (1-\pi(X))\\
&+  [E(Y_{0M_1}\mid X) - E(Y_{0M_0}\mid X)]^2 \pi(X)\bigg],
\end{align*}
which demonstrates that it is fully due to the indirect between-hospital effect on the length of stay via the mediator.
\end{description}

\subsection{Hypothetical assignment mechanism}\label{Hypothetical}

Because the between-hospital variance in Equation (\ref{3-way}) was derived for the observed marginal variance of the outcome, the causal mediation decomposition depends on the hospital assignment probabilities $e(z; X)$. However, we can also consider decomposition under hypothetical ``randomized'' assignment mechanism, where for example each hospital treats similar kind of patient population, and/or similar patient volume.  Let $A$ represent a random draw from such a hypothetical assignment mechanism with probabilities $\tilde{e} (a; X) = P(A = a\mid X)$, chosen so that the causal assumptions in Section \ref{Notation and assumptions} are satisfied. The variance decomposition can then be written for the marginal variance $V[Y_A]$, which will be of the same form as \eqref{3-way}, but with the assignment probabilities $e(z; X)$ replaced with $\tilde{e} (a; X)$. Choosing for example $\tilde{e} (a; X) = P(Z = a\mid X)$ would give the same decomposition, while choosing $\tilde{e} (a; X) = 1/q$ would correspond to a mechanism where each hospital treats similar patient population of the same size. In the latter case the between-hospital variance component would have a causal interpretation similar to the model reliance metric considered in a machine learning context by \citet{fisher2019all}. Now, the causal mediation decomposition for the between-hospital variance can be written as
\begin{align}\label{mediation_hypothetical}
\MoveEqLeft  \sum_{a}\bigg[E(Y_a \mid X)  - \sum_{a^*} E(Y_{a^*} \mid X) \tilde{e}(a^*; X) \bigg]^2 \tilde{e}(a; X) \nonumber\\
&= \sum_{a}\bigg[E(Y_{aM_a} \mid X)  - \sum_{a^*} E(Y_{aM_{a^*}} \mid X) \tilde{e} (a^*; X) \bigg]^2 \tilde{e} (a; X) \nonumber\\
&\quad+ \sum_{a} \bigg[ \sum_{a^*} E(Y_{aM_{a^*}} \mid X) \tilde{e} (a^*; X) - \sum_{a^*} E(Y_{a^*M_{a^*}} \mid X)\tilde{e} (a^*; X) \bigg]^2 \tilde{e} (a; X) \nonumber\\
&\quad+ \sum_{a}2 \bigg[E(Y_{aM_a} \mid X)  - \sum_{a^*} E(Y_{aM_{a^*}} \mid X) \tilde{e} (a^*; X) \bigg] \nonumber\\
&\quad\quad\times \bigg[ \sum_{a^*} E(Y_{aM_{a^*}} \mid X) \tilde{e} (a^*; X) - \sum_{a^*} E(Y_{a^*M_{a^*}} \mid X) \tilde{e} (a^*; X) \bigg] \tilde{e} (a; X).
\end{align}

To understand the connetion of the $R^2$ type mediation decomposition in Equation (\ref{R2_type}), it is helpful to consider a special case of two hospitals with the setting the
assignment probabilities as $\tilde{e} (a; X)=1/2$. Now, the between-hospital variation becomes
\begin{align*}
\frac{1}{4}\bigg[E(Y_0\mid X)-E(Y_1\mid X) \bigg]^2.
\end{align*}
The indirect effect term becomes
\begin{align*}
\frac{1}{8}\bigg[ [E(Y_{1M_1}\mid X)- E(Y_{1M_0}\mid X)]^2 +  [E(Y_{0M_1}\mid X)- E(Y_{0M_0}\mid X)]^2\bigg],
\end{align*}
the direct effect term becomes
\begin{align*}
\frac{1}{8}\bigg[ [E(Y_{1M_0}\mid X)- E(Y_{0M_0}\mid X)]^2 +  [E(Y_{1M_1}\mid X)- E(Y_{0M_1}\mid X)]^2\bigg],
\end{align*}
and the covariance term between the indirect and direct effects becomes
\begin{align*}
2\times\frac{1}{8}\bigg[& [E(Y_{1M_1}\mid X)- E(Y_{1M_0}\mid X)] [E(Y_{1M_0}\mid X)- E(Y_{0M_0}\mid X)]\\
&\quad+ [E(Y_{0M_0}\mid X)- E(Y_{0M_1}\mid X)] [E(Y_{0M_1}\mid X)- E(Y_{1M_1}\mid X)]\bigg].
\end{align*}
Under the causal assumptions of Section \ref{Notation and assumptions}, and the three linear models in \ref{3linear},
\begin{align*}
E(Y_0\mid X=x)&=E[Y_0\mid A=0,X=x]\\
&=E[Y_0\mid Z=0,X=x]\\
&=E[Y\mid Z=0,X=x]\\
&=\alpha_1+\gamma_1x,
\end{align*}
and $E(Y_1\mid X=x)=\alpha_1+\beta_{1}+\gamma_1x.$ Also, we have
\begin{align*}
E(Y_{1M_{0}} \mid X=x)&=\int_m E[Y_{1m} \mid X=x]f_{M_{0}}(m \mid X=x) \,\textrm dm \nonumber\\
&=\int_m E[Y \mid Z=1, M=m, X=x]f_M(m \mid Z=0,X=x) \,\textrm dm\\
&=\int_m (\alpha_2+\beta_{2}+\beta_{3}m+\gamma_2x)f_M(m \mid Z=0,X=x) \,\textrm dm\\
&=\alpha_2+\beta_{2}+\beta_3E[M\mid Z=0, X=x]+\gamma_2x\\
&=\alpha_2+\beta_{2}+\beta_3(\alpha_3+\gamma_3x)+\gamma_2x.
\end{align*}
Similarly, we have
\begin{align*}
E(Y_{0M_{0}} \mid X=x)&=\alpha_2+\beta_3(\alpha_3+\gamma_3x)+\gamma_2x, \\
E(Y_{0M_{1}} \mid X=x)&=\alpha_2+\beta_3(\alpha_3+\beta_4+\gamma_3x)+\gamma_2x, \\
E(Y_{1M_{1}} \mid X=x)&=\alpha_2+\beta_2+\beta_3(\alpha_3+\beta_4+\gamma_3x)+\gamma_2x.
\end{align*}
Substituting the above results into the components of causal mediation decomposition, the total between-hospital variance, the total effect becomes $\frac{1}{4}\beta_1^2$ and the three terms on the right hand side become $\frac{1}{4}(\beta_3\beta_4)^2,\frac{1}{4}\beta_2^2$, and $2\times\frac{1}{4}\beta_2\beta_3\beta_4$. Hence, we have
$\beta_1^2=\beta_2^2+(\beta_3\beta_4)^2+2\beta_2\beta_3\beta_4$, which is same as equation (\ref{R2_type}). This shows that the decomposition discussed by \cite{de2012r} in the linear structural equation modeling framework is a special case of the our decomposition \eqref{mediation} expressed in terms of potential outcomes, obtained in the case of identity link and in the absence of exposure-mediation interaction.

\section{Estimators}\label{section:estimation}

\subsection{Point estimation}\label{Point estimation}

We note that due to the counterfactual consistency, we have that $E[Y_{zM_{z}} \mid X] = E[Y_{z} \mid X]$, $E[Y_{zM_{Z}} \mid X] = E[Y_{zM} \mid X]$ and $E[Y_{ZM_{Z}} \mid X] = E[Y \mid X]$, and thus estimation of the decomposition \eqref{mediation} could in principle be based on modeling the quantities $E[Y \mid Z, X]$, $E[Y \mid Z, M, X]$, $E[M \mid X]$ and $E[Y \mid X]$. However, with a view of estimating also decompositions of the type \eqref{mediation_hypothetical}, and to enable estimation of uncertainty based on factorization of the likelihood (Section \ref{Variance_estimation}), we propose an estimation approach  based on fitting a hospital, mediator and case-mix conditional outcome model for $E[Y \mid Z, M, X]$, a hospital and case-mix conditional mediator model for $E[M \mid Z, X]$, and a case-mix conditional hospital assignment model for $P(Z \mid X)$, under the causal assumptions listed in Section \ref{Notation and assumptions}. 

As an example we consider a conditional structural model for the outcome which under the strong ignorability assumption can be specified as
\begin{align}\label{mixedmodel}
E[Y_{zm} \mid X=x; \theta_1]&=E[Y \mid Z=z, M=m, X=x; \theta_1] \nonumber \\
&= g^{-1}\left(\beta_0+\beta_{z}+\beta_{q+1}m + \beta_{(q+2):(q+2+p)}^{'}x \right),
\end{align}
where $\beta_{(q+2):(q+2+p)} = (\beta_{q+2}, \ldots, \beta_{q+2+p})$, $\theta_1 = (\beta_0, \beta_1,\ldots, \beta_q, \beta_{q+1}, \beta_{q+2},\ldots,\beta_{q+2+p})$, and $g$ is an arbitrary link function. The hospital level intercept terms $\beta_{z}, z\in\{1,...,q\}$ are either taken to be fixed effects, with $\beta_1 = 0$, or follow $\beta_{z} \sim N(0,\tau_1^2)$ in a random intercept model, in which case \eqref{mixedmodel} still applies conditional on the true values of the random effects. $\beta_{q+1}$ represents the mediator effects on the outcome. Interaction terms could be added to the model as needed.

Under strong ignorability, a conditional structural model for the mediator can be specified as
\begin{align}\label{mediatormodel}
E[M_z \mid X=x;\theta_2]&= E[M \mid Z=z,X=x;\theta_2] \nonumber \\
&=h^{-1}\left(\gamma_0+\gamma_{z} +\gamma_{(q+1):(q+1+p)}^{'}x \right),
\end{align}
where $\gamma_{(q+1):(q+1+p)} = (\gamma_{q+1}, \ldots, \gamma_{q+1+p})$, $\theta_2 = (\gamma_0, \gamma_{1}, \ldots, \gamma_{q}, \gamma_{q+1}, \ldots, \gamma_{q+1+p})$, and $h$ is an arbitrary link function. The hospital level intercept terms $\gamma_{z}, z\in\{1,...,q\}$ are either fixed, with $\gamma_1 = 0$, or follow $\gamma_{z} \sim N(0,\tau_2^2)$  in random intercept model. 

With a binary mediator, using the link $h(.) = \textrm{logit}(.)$, combining \eqref{mixedmodel} and  \eqref{mediatormodel}, the covariate conditional expected potential outcomes at different levels of $z$ (direct) and $z^*$ (indirect) can be calculated as
\begin{align}\label{equation:estimator}
E(Y_{zM_{z^*}} \mid X=x, \theta_1,\theta_2)&=E_{M_{z^*}\mid X}\{ E(Y_{zM_{z^*}} \mid M_{z^*}, X=x, \theta_1,\theta_2)\} \nonumber\\
&=\sum_{m=0}^{1}E[Y_{zm} \mid X=x; \theta_1]P[M_{z^*}=m \mid X=x;\theta_2]  \nonumber\\
&= \sum_{m=0}^{1} E[Y \mid Z=z, M=m, X=x; \theta_1] \nonumber\\
&\quad\quad \times P[M=m \mid Z=z^*,X=x;\theta_2] \nonumber\\
&= g^{-1}\left(\beta_0+\beta_{z}+\beta_{q+1}m + \beta_{(q+2):(q+2+p)}^{'}x \right) \nonumber\\
&\quad\times\text{expit}\left(\gamma_0+\gamma_{z^*} +\gamma_{(q+1):(q+1+p)}^{'}x \right) \nonumber\\
&\quad +g^{-1}\left(\beta_0+\beta_{z}+ \beta_{(q+2):(q+2+p)}^{'}x \right) \nonumber\\
&\quad\quad\times\left[1-\text{expit}\left(\gamma_0+\gamma_{z^*} +\gamma_{(q+1):(q+1+p)}^{'}x \right)\right],
\end{align}
where the second and third equalities follows from the third and fourth causal assumptions.

For a continuous mediator we have instead
\begin{align}\label{equation:mediator}
E(Y_{zM_{z^*}} \mid X=x, \theta_1,\theta_2)&=\int_m E[Y_{zm} \mid X=x; \theta_1]f_{M_{z^*}}(m \mid X=x;\theta_2) \,\textrm dm \nonumber\\
&=\int_m E[Y \mid Z=z, M=m, X=x; \theta_1] \nonumber\\
&\quad\quad\times f_M(m \mid Z=z^*,X=x;\theta_2) \,\textrm dm,
\end{align}
where $f_M(m \mid Z=z^*,X=x;\theta_2)$ is the density function of mediator based on a distributional assumption, such as normally distributed residuals in a linear regression model.

For the hospital assignment mechanism, we fit a multinomial logistic regression model
\begin{equation}\label{equation:multinomial}
e(z, x; \theta_3) =
\begin{cases}
\frac{1}{1+\sum_{z'=2}^q \exp(\psi_{z'}+\phi_{z'}'x)} & z=1\\
\frac{ \exp(\psi_{z}+\phi_z'X)}{1+\sum_{z'=2}^q \exp(\psi_{z'}+\phi_{z'}'x)}& z\neq1,
\end{cases}
\end{equation}
where $\theta_3 = (\psi_2, \ldots, \psi_q, \phi_2, \ldots, \phi_q)$.

Further, we denote $\omega_0(\theta_1,\theta_2,\theta_3), \omega_1(\theta_1,\theta_2,\theta_3), \omega_2(\theta_1,\theta_2,\theta_3)$ and $\omega_3(\theta_1,\theta_2,\theta_3)$ as the parametrized versions of the total between-hospital variance and the three additive terms in the decomposition \eqref{mediation}. Hence, we have $\omega_0(\theta_1,\theta_2,\theta_3)= \omega_1(\theta_1,\theta_2,\theta_3)+\omega_2(\theta_1,\theta_2,\theta_3)+\omega_3(\theta_1,\theta_2,\theta_3)$.  We introduce the shorthand notations $\mu(z,m,x;\theta_1) = E[Y_i \mid Z_i=z, M_i=m, X_i=x; \theta_1]$ and $\eta(m; z,x;\theta_2) = P[M_i=m \mid Z_i=z, X_i=x; \theta_2]$. 

The between-hospital variance can be estimated by
\begin{align*}
\omega_0(\hat\theta_1,\hat\theta_2,\hat\theta_3)& = \frac{1}{n} \sum_{i=1}^n \bigg\{ \sum_{z}\bigg[ \sum_{m=0}^1 \mu(z,m,x_i;\hat\theta_1)\eta(m,z,x_i;\hat\theta_2) \\
&\quad - \sum_{z^*} \sum_{m=0}^1 \mu(z^*,m,x_i;\hat\theta_1)\eta(m,z^*,x_i;\hat\theta_2) e(z^*, x_i; \hat\theta_3)  \bigg]^2  e(z, x_i; \hat\theta_3)  \bigg\}.
\end{align*}
The indirect effect term can be estimated by
\begin{align*}
\omega_1(\hat\theta_1,\hat\theta_2,\hat\theta_3)&= \frac{1}{n} \sum_{i=1}^n \bigg\{  \sum_{z}\bigg[\sum_{m=0}^1 \mu(z,m,x_i;\hat\theta_1)\eta(m,z,x_i;\hat\theta_2) \\
&\quad - \sum_{z^*} \sum_{m=0}^1 \mu(z,m,x_i;\hat\theta_1)\eta(m,z^*,x_i;\hat\theta_2)  e(z^*, x_i; \hat\theta_3) \bigg]^2  e(z, x_i; \hat\theta_3) \bigg\}.
\end{align*}
The direct effect term can be estimated by
\begin{align*}
\omega_2(\hat\theta_1,\hat\theta_2,\hat\theta_3)&= \frac{1}{n} \sum_{i=1}^n \bigg\{  \sum_{z} \bigg[ \sum_{z^*} \sum_{m=0}^1 \mu(z,m,x_i;\hat\theta_1)\eta(m, z^*,x_i;\hat\theta_2)  e(z^*, x_i; \hat\theta_3)\\
&\quad - \sum_{z^*} \sum_{m=0}^1 \mu(z^*,m,x_i;\hat\theta_1)\eta(m,z^*,x;\hat\theta_2) e(z^*, x_i; \hat\theta_3) \bigg]^2  e(z, x_i; \hat\theta_3)  \bigg\}.
\end{align*}
The covariance term between the indirect and direct effects can be estimated as
\begin{align*}
\omega_3(\hat\theta_1,\hat\theta_2,\hat\theta_3)=\omega_0(\hat\theta_1,\hat\theta_2,\hat\theta_3)-\omega_1(\hat\theta_1,\hat\theta_2,\hat\theta_3)-\omega_2(\hat\theta_1,\hat\theta_2,\hat\theta_3).
\end{align*}
By the continuous mapping theory and the law of large numbers, if the parametric models can be consistently estimated such that $\hat\theta_j  \stackrel{p}{\rightarrow} \theta_j$ for $j \in \{1,2,3\}$, then the variance component estimators $\omega_s(\hat\theta_1,\hat\theta_2,\hat\theta_3)$ for $s \in \{0,1,2,3\}$ will be also consistent. We will investigate their finite sample behaviour in the simulation study in Section \ref{section:simulation}.

\subsection{Variance estimation}\label{Variance_estimation}
The estimators proposed in section \ref{Point estimation} are fully model-based. We can evaluate their uncertainty via approximate Bayesian inference by factorization of the likelihood. We can draw the samples from the joint posterior distribution of $\theta_j$ for $j \in \{1,2,3\}$,
\begin{align*}
\MoveEqLeft f(\theta_1,\theta_2,\theta_3 \mid \mathbf{Y}, \mathbf{Z}, \mathbf{M}, \mathbf{X}) \\
&= \frac{f(\mathbf{Y}, \mathbf{Z}, \mathbf{M}  \mid \mathbf{X}, \theta_1,\theta_2, \theta_3) f(\theta_1 \mid \mathbf{X})  f(\theta_2 \mid \mathbf{X}) f(\theta_3 \mid \mathbf{X})}{f(\mathbf{Y}, \mathbf{Z}, \mathbf{M} \mid \mathbf{X})} \\
&= \frac{f(\mathbf{Y} \mid  \mathbf{Z},  \mathbf{M}, \mathbf{X}, \theta_1) f(\theta_1 \mid \mathbf{X})}{f(\mathbf{Y} \mid \mathbf{Z}, \mathbf{M}, \mathbf{X})} 
\frac{f(\mathbf{M} \mid  \mathbf{Z}, \mathbf{X}, \theta_2) f(\theta_2 \mid \mathbf{X})}{f(\mathbf{M} \mid \mathbf{Z}, \mathbf{X})} 
\frac{f(\mathbf{Z} \mid \mathbf{X}, \theta_3) f(\theta_3 \mid \mathbf{X})}{f(\mathbf{Z} \mid \mathbf{X})} \\
&= f(\theta_1 \mid \mathbf{Y}, \mathbf{Z}, \mathbf{M},\mathbf{X}) f(\theta_2 \mid \mathbf{Z}, \mathbf{M},\mathbf{X})f(\theta_3 \mid \mathbf{Z}, \mathbf{X}),
\end{align*}
where we assumed that $\theta_1$, $\theta_2$ and $\theta_3$ are a priori independent given $X$. If the parameters are estimated through Markov chain Monte Carlo, the resulting posterior samples can be directly used for inferences on the variance components. Alternatively, the posterior distributions can be approximated asymptotically, in the sense of Bernstein-von Mises theorem \citep[][Chapter 10]{van2000asymptotic}, as normal distributions centered at the maximum likelihood estimator and variance covariance matrix given by the inverse Fisher information. Bootstrap-based approximation of posterior distributions is also possible \citep{newton1994approximate}. We sample the parameter $\theta_j$ ($j \in \{1,2,3\}$) from their approximate posterior distribution, and recalculate the variance components $\omega_k(\theta_1,\theta_2,\theta_3)$  ($k \in \{0,1,2,3\}$) for each draw. For the outcome model parameter $\theta_1$ and the mediator model parameter $\theta_2$, we use the parametric bootstrapping to approximate their posterior distribution. We resample the outcomes/mediators from the fitted models, refit the models, and then calculate the new fitted values. For the hospital assignment parameters $\theta_3$, we use the normal approximation to sample them from the multivariate normal distributions $\theta_3\sim MVN(\hat{\theta}_3,V(\hat{\theta}_3))$, where $\hat\theta_3$ are the maximum likelihood estimators and $V(\hat\theta_3)$ are their asymptotic variance-covariance matrix from the multinomial logistic model fit.

\section{Simulation study}\label{section:simulation}

\subsection{Generating mechanism}\label{section:generation}

We demonstrated the performance of the estimators via a simulation study, using a data generating mechnism similar to the one in Figure \ref{figure:dag}. The objectives for the simulation study were to (a) study the finite sample properties of the proposed point estimators for both continuous and binary outcomes under different $n$ (total number of patients) and $q$ (numer of hospitals), (b) to check the performance of the point estimators in the cases zero and non-zero covariance between the indirect and direct effects, and (c) to check the performance of the point estimators when in the absence of the mediator effects on the outcomes directly (absence the arrow $M \rightarrow Y$ in Figure \ref{figure:dag}). The asymptotic behavior of the estimators was studied by varying the total number of hospitals $m$ and the total number of patients $n$. To begin with, we generated two patient case-mix factors, $X_1\sim N(0,1)$ and $X_2\sim \textrm{Bernoulli}(0.5)$. The hospital ($Z$) assignment was generated based on multinomial logistic model with the two case-mix factors, where the hospital-specific intercepts were generated from $N(0,0.25)$ and coefficients were generated from $N(0,0.5)$, specifying the parameter vector $\theta_3$. A latent continuous mediator $M_z$ was generated from a mean structure model
\begin{equation}\label{generator1}
E[M_z\mid X]= \gamma_{z} + X_{1} + 1.5 X_{2},
\end{equation}
where $\gamma_{z}$ is the effect of hospital $z$ on the mediator. The outcomes were simulated as $M_z = E[M_z\mid X] + \varepsilon$, where $\varepsilon_z \sim \textrm{Logistic}(0,1)$. The binary mediator $M_z^*$ was generated via dichotomizing the continuous mediator as $M_z^*= \mathbf 1_{\{M_z \ge 0\}}$. Only the dichotomized version of the mediator was used for the simulation studies. The continuous outcomes were generated from the mean structure model
\begin{equation}\label{generator2}
E[Y_{zm} \mid X]= \beta_{z} +\beta_{q+1} m + X_{1} + 2 X_{2}
\end{equation}
where $\beta_{z}$ is the effect of hospital $z$ on the outcome, and $\beta_{q+1}$ is the mediator effect on the outcome. The outcomes were simulated as $Y_{zm} = E[Y_{zm} \mid X] + \xi$, where $\xi \sim \textrm {Logistic}(0,1)$, with the observed outcome generated as $Y = Y_{ZM_Z^*}$ given the observed values $Z$ and $M^* = M_Z^*$ of the hospital assignment and binary mediator, respectively. The binary outcomes $Y_{zm}^*$ were generated via dichotomizing the continuous outcome as $Y_{zm}^* = \mathbf 1_{\{Y_{zm} \ge 0\}}$. 

The hospital effects on the mediator and the outcome were generated from
\begin{equation}\label{generator3}
\begin{bmatrix}
     \gamma_{z}\\
    \beta_{z}\\
\end{bmatrix}
\sim MVN 
\left(
\begin{bmatrix}
     0\\
   	0
\end{bmatrix},
\begin{bmatrix}
     4 &\sigma\\
   	\sigma&4
\end{bmatrix}
\right)
\end{equation} 
for $z = 1, \ldots, q$, where we chose $\sigma=0$ to produce a scenario where the indirect and direct effects are uncorrelated, and $\sigma=2$ for a scenario where these were correlated. The mediator effect on the outcome $\beta_{q+1}$ was varied from 7, 4.7, and 0 for the continuous outcome, binary outcome, and absence of mediator effects on the outcome scenarios, respectively. 

In the simulation study, we considered sampling variation generated by independent samples of patients being treated in the same hospitals (e.g. over a given time period). Thus, the hospital coefficients were sampled only once for each simulation scenario and then fixed across the replications, while an independent sample of patients was generated in each replication. This corresponds to the administrative data setting of Section \ref{section:illustration} where all the hospitals of the administrative region are observed, along with patients in a given time period, with the inferences then corresponding to an unobserved `long run' performance of these hospitals. For the same reasons, we only considered random effects models as means for estimation, rather than data generating mechanisms, as we are not considering random samples of hospitals from a population of hospitals. Random effect models were fitted to data simulated from the fixed effect generating mechanism using functions in the R package lme4 \citep{lme4}, with the resulting empirical Bayes predictions \citep[e.g.][Chapter 7]{skrondal2004generalized} used as estimates for the hospital effects in equations \eqref{mixedmodel} and \eqref{mediatormodel}. All the other parameters were substituted with their maximum likelihood estimates.

\subsection{Computation of true values of the variance components based on the generating mechanism}\label{section:true}

The parameters specified in the generating distributions specify the true values of the variance components of interest, but the latter are fairly complicated functions of the former. To carify this connection, we derived the formulas for the three variance components as functions of the generating mechanism in the binary mediator and continuous outcome case. Because the outcome model is a linear function of the mediator, some simplifications are possible. Firstly, we substituted the models \eqref{generator1} and \eqref{generator2} into Equation \eqref{equation:estimator}, and summing over the mediator levels, the true value of the covariate conditional expected potential outcome at different levels of $z$ (direct) and $z^*$ (indirect) is given by
\begin{align}\label{equation_true_1}
E(Y_{zM_{z^*},i} \mid x_i)=\beta_z+\beta_{q+1} \times \textrm{expit}(\gamma_{z^*}+ x_{1,i} + 1.5 x_{2,i})+ x_{1,i} + 2 x_{2,i}.
\end{align}
Then, we substituted \eqref{equation_true_1} into the equation \eqref{mediation}. Due to most of terms in the outcome model cancelling out from the differences, the true values of the indirect effect, direct effect, and covariance terms given the simulated covariate distribution are given by
\begin{align*}
\frac{\beta_{q+1}^2}{n}\sum_{i=1}^n \bigg \{\sum_{z}\Big[& \textrm{expit}(\gamma_{z} + x_{1,i} + 1.5 x_{2,i})\\
&-\sum_{z^*}\textrm{expit}(\gamma_{z^*} + x_{1,i} + 1.5 x_{2,i}) \bar{e} (z^*,x_i; \theta_3)\Big]^2 e(z,x_i; \theta_3)\bigg\},
\end{align*}
\begin{align*}
& \frac{1}{n}\sum_{i=1}^n \bigg \{\sum_{z}\Big[\beta_{z}-\sum_{z^*}\beta_{z^*} e(z^*,x_i; \theta_3)\Big]^2 e(z,x_i; \theta_3)\bigg\},
\end{align*}
and 
\begin{align*}
& \frac{2\beta_{q+1}}{n}\sum_{i=1}^n \bigg \{\sum_{z}\Big[\textrm{expit}(\gamma_{z} + x_{1,i} + 1.5 x_{2,i})-\\
&\sum_{z^*}\textrm{expit}(\gamma_{z^*} + x_{1,i} + 1.5 x_{2,i}) e(z^*,x_i; \theta_3)\Big] \Big[\beta_{z}-\sum_{z^*}\beta_{z^*} 
e(z^*,x_i)\Big]e(z,x_i; \theta_3)\bigg\},
\end{align*}
respectively. Because the hospital effects and assignment probabilities were recreated for each simulation scenario with different numbers of hospitals, the true values of the components varied across the simulation scenarios. As noted at the end of the previous section, the hospital effects were fixed in the data generating mechanism, and thus the above expressions corresponding to fixed effects models were also used as the true values when comparing to estimators given by random effects models. We note that if all the hospitals had the same assignment probability $1/q$ (no case-mix differences between the hospitals), or we were considering the decomposition under such a hypothetical assignment mechanism, the direct effect term would converge to $V(\beta_z)=4$ in \eqref{generator3} when $q \rightarrow \infty$, which is expected as under the linear outcome model without exposure-mediator interaction, the hospital effects $\beta_z$ directly represent the direct effects.

\subsection{Results: continuous outcomes}\label{simulation_continuous}

\begin{figure}[!ht]
\centering
\includegraphics[width=0.92\textwidth]{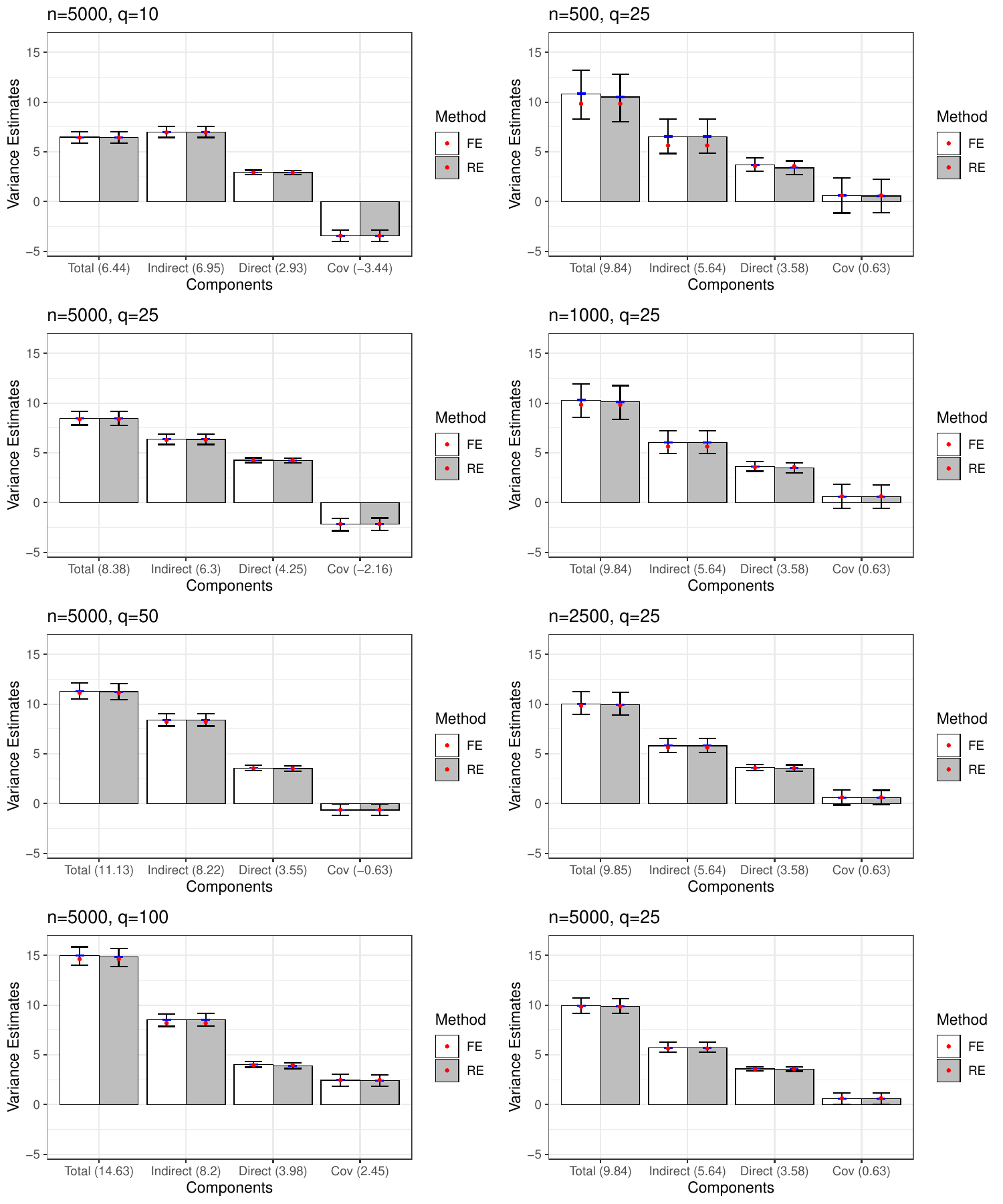}
\caption{Simulated sampling distribution means for the variance components under fixed and random effect models for the continuous outcomes under different combinations of $n$ (total number of patients) and $q$ (total number of hospitals), based on 1000 replications. The dots indicate the true values. The wide error bar represents the $95\%$ quantile interval of the sampling distribution. The  narrow error bar represents the $95\%$ confidence interval for the mean, reflecting the Monte Carlo error. The right hand panels represent the same simulation scenario (are comparable), while the left hand panels represent different scenarios.}
\label{Plot1}
\end{figure}

Figure \ref{Plot1} shows the simulated sampling distribution means for the total between-hospital variance, the indirect effect term, the direct effect term, and the covariance term for the continuous outcomes, under different combinations of $n$ (total number of patients), $q$ (number of hospitals), and with $\sigma=0$, based on 1000 replications. The white bars and gray bars represent the estimates from the fixed- and random-effect models. The dots indicate the true value, and the error bars with wide caps show the $95\%$ quantile interval of the sampling distribution. The $95\%$ confidence interval for the mean is represented by the error bars with narrow caps, reflecting the Monte Carlo error in the estimated mean of the sampling distribution. We note that the Monte Carlo error is so small relative to the scale of the plot that the latter bars often appear as a single line.

From the plot, we observe that the total between-hospital variance, the indirect effect term, the direct effect term, and the covariance term are well estimated with both fixed and random effect models when the per hospital number of patients is sufficiently large. However, some positive small sample bias is present in the indirect effect estimates when this is small, in particular in the scenario $n=500$, $q=25$, when there is on average only 20 patients per hospital. We note that the different values of $q$ all represent different scenarios of the data generating mechanism, so the true values of the components are different on the left hand column, but are the same in the right hand column (can be compared across rows). Also, while the hospital effects for the outcome and the mediator were simulated from a joint distribution with zero correlation, the sample correlation of these in the simulation scenarios may be non-zero, which is reflected by the different true values for the covariance term. 

\begin{figure}[!ht]
        \centering
        \subfloat[]{\label{Plot2}\includegraphics[width=0.5\textwidth]{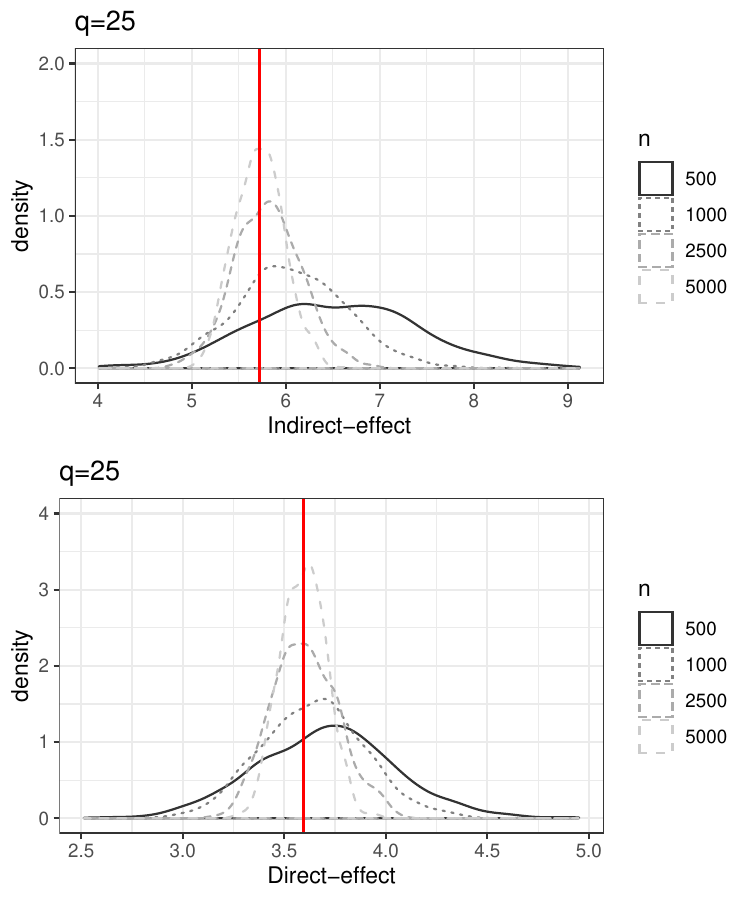}}
        \subfloat[]{\label{Plot3}\includegraphics[width=0.5\textwidth]{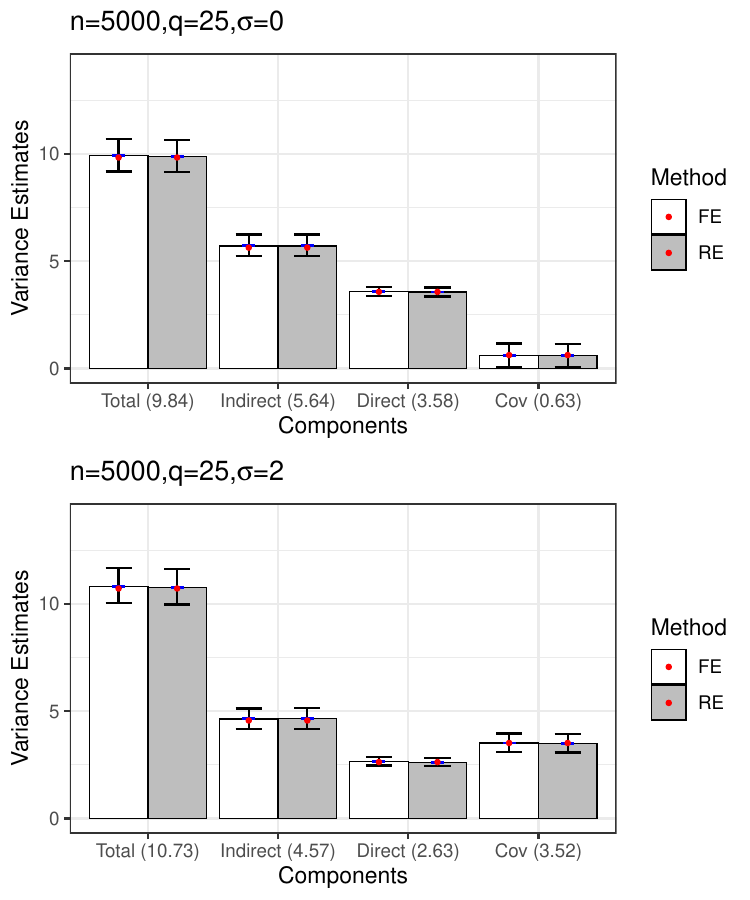}}
        \caption{Panel (a): Density plots for the simulated sampling distributions of the indirect and direct effect components with a continuous outcome, based on 1000 replications. The red vertical line indicates the true value. Panel (b): Simulated sampling distribution means for the decomposition of between-hospital variance in the continuous outcomes, based on 1000 replications. The top and bottom plots show the covariance between the indirect and direct effects is equal to zero and non-zero in the data generation \eqref{generator3} ($\sigma=0$ and $\sigma=2$). The dots indicate the true values. The wide error bar represents the $95\%$ quantile interval of the sampling distribution. The  narrow error bar represents the $95\%$ confidence interval for the mean, reflecting the Monte Carlo error.}
\end{figure}

\begin{figure}[!ht]
\centering
\includegraphics[width=\textwidth]{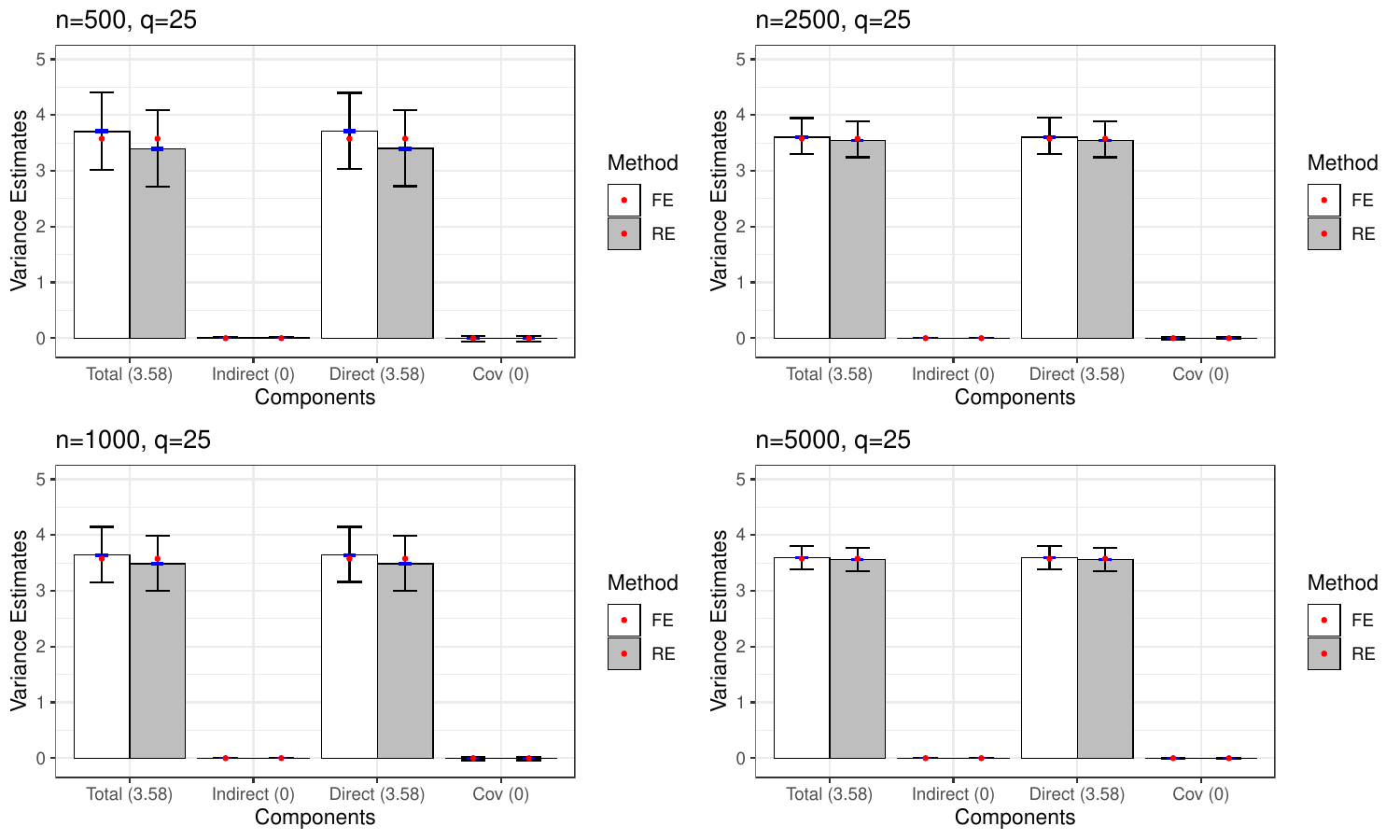}
\caption{Simulated sampling distribution means for the variance components under the fixed- and random-effect models for the continuous outcomes under the different total number of patients $n$, based on 1000 replications. The direct mediator effect on the outcome was absent (absence of the arrow $M \rightarrow Y$ in Figure \ref{figure:dag}) in data generation. The dots indicate the true values. The wide error bar represents the $95\%$ quantile interval of the sampling distribution. The  narrow error bar represents the $95\%$ confidence interval for the mean, reflecting the Monte Carlo error.}
\label{Plot4}
\end{figure}

Figure \ref{Plot2} shows the density plots for the simulated sampling distributions of the indirect and direct effect variance components with a continuous outcome, based on 1000 replications. We observe that the densities are fairly symmetric, and become more concentrated when $n$ increases, reflecting the consistency of the estimators. Figure \ref{Plot3} shows simulated sampling distribution means for the decomposition of between-hospital variance in the continuous outcomes between scenarios where $\sigma=0$ and $\sigma=2$, based on 1000 replications. We observe that the covariance term estimates are close to 0 when there is no correlation between the direct and indirect effects in the data generating mechanism, and close to the true value of 4 when these are correlated. Figure \ref{Plot4} shows the simulated sampling distribution means for the variance components under the fixed- and random-effect models for the continuous outcomes under the different total number of patients $n$ in the scenario with $\beta_{q+1}=0$ (absence of the arrow $M \rightarrow Y$ in Figure \ref{figure:dag}), based on 1000 replications. From the plot, we can observe that the indirect effect term and the covariance term estimate to close to 0, and the total between-hospital variance is fully contributed by the direct effect.

The computation of the variance estimator proposed in Section \ref{Variance_estimation} was slow and thus we did not implement it across all of the scenarios. However, to investigate its performance in a scenario that most closely resembles the real data application in Section \ref{section:illustration} (continuous outcome, estimation based on mixed effect models, $n=5000$, $q=50$), we ran $50$ additional simulation rounds and compared the resulting Monte Carlo standard deviations of the point estimates to the average standard errors obtained from 50 draws of the approximate Bayesian procedure for each simulation round. The average standard errors of the total between-hospital variance, the indirect effect term, the direct effect term, and the covariance term were $0.4359$, $0.3428$, $0.1370$, and $0.3289$, respectively, compared to the corresponding Monte Carlo standard deviations $0.4312$, $ 0.3423$, $0.1372$, and $0.3295$. As the results were very close, together with the small bias of the point estimators and a symmetric sampling distribution of these, with large enough sample size normal approximation based intervals could be calculated using the variance estimates. However, with a small sample we recommend using more draws in the approximate Bayesian procedure and calculating quantile-based intervals.

\subsection{Results: binary outcomes}\label{simulation_binary}

\begin{figure}[!ht]
\centering
\includegraphics[width=0.92\textwidth]{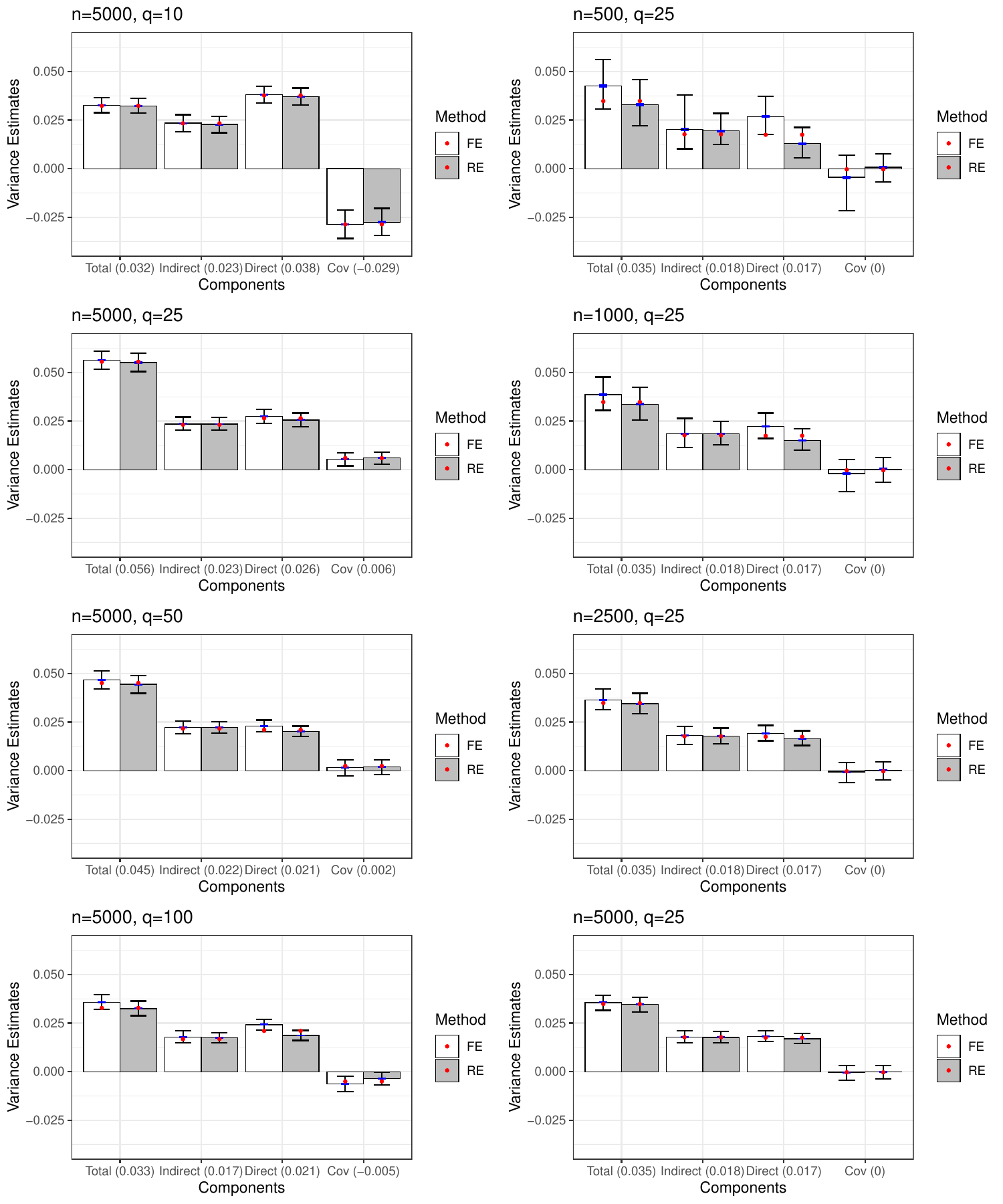}
\caption{Simulated sampling distribution means for the variance components under the fixed- and random-effect models for the binary outcomes under different combinations of $n$ (total number of patients) and $q$ (total number of hospitals), based on 1000 replications. The red dots indicate the true variances. The $95\%$ quantile interval of the sampling distribution is represented by the black error bar. The $95\%$ confidence interval for the mean is represented by the blue error bar, reflecting the Monte Carlo error in the estimated mean of the sampling distribution. The right hand panels represent the same simulation scenario (are comparable), while the left hand panels represent different scenarios.}
\label{Plot5}
\end{figure}

\begin{figure}[!ht]
        \centering
        \subfloat[]{\label{Plot6}\includegraphics[width=0.5\textwidth]{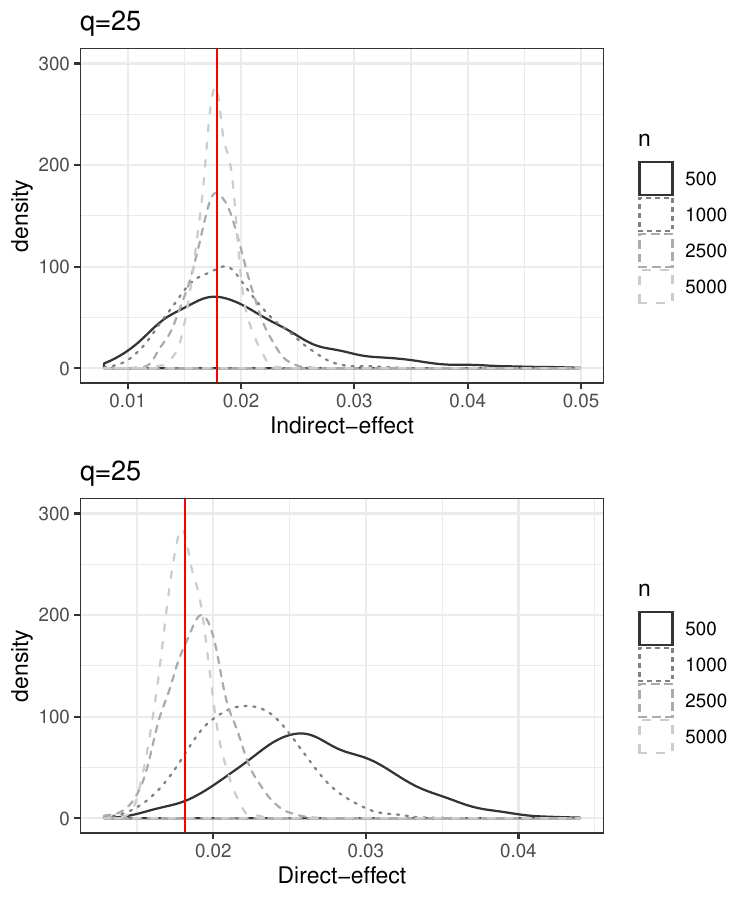}}
        \subfloat[]{\label{Plot7}\includegraphics[width=0.5\textwidth]{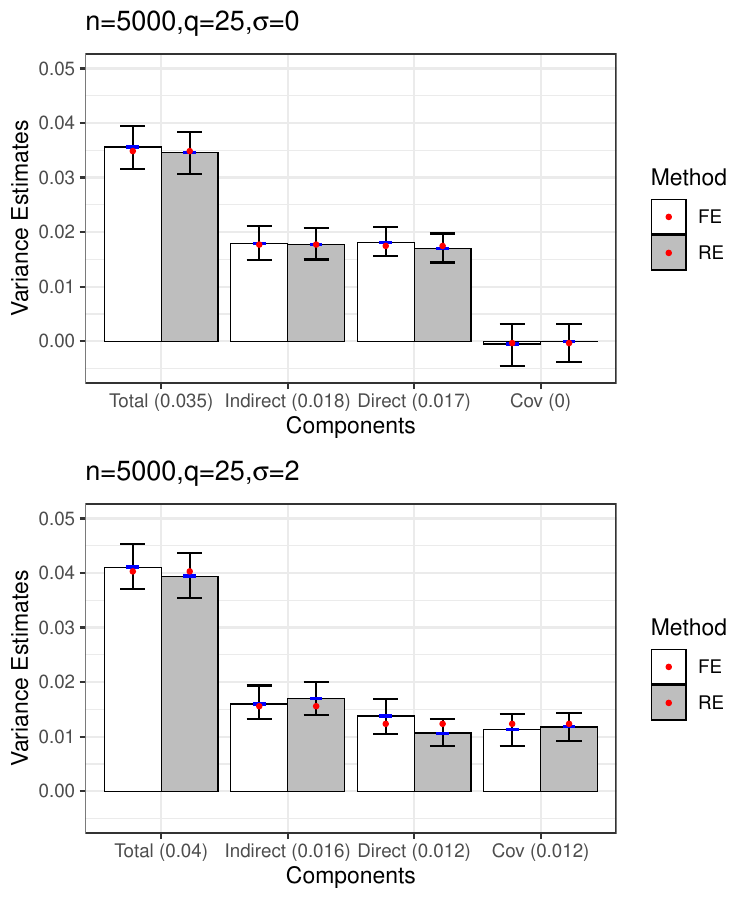}}
        \caption{Panel (a): Density plots for the simulated sampling distributions of the indirect and direct effect variance components with a binary outcome, based on 1000 replications. The red vertical line indicates the true value. Panel (b): Simulated sampling distribution means for the decomposition of between-hospital variance in the binary outcomes, based on 1000 replications. The top and bottom plots show the covariance between the indirect and direct effects is equal to zero and non-zero in the data generation \eqref{generator3} ($\sigma=0$ and $\sigma=2$). The red dots indicate the true variances. The $95\%$ quantile interval of the sampling distribution is represented by the black error bar. The $95\%$ confidence interval for the mean is represented by the blue error bar, reflecting the Monte Carlo error in the estimated mean of the sampling distribution.}
\end{figure}

\begin{figure}[!ht]
\centering
\includegraphics[width=\textwidth]{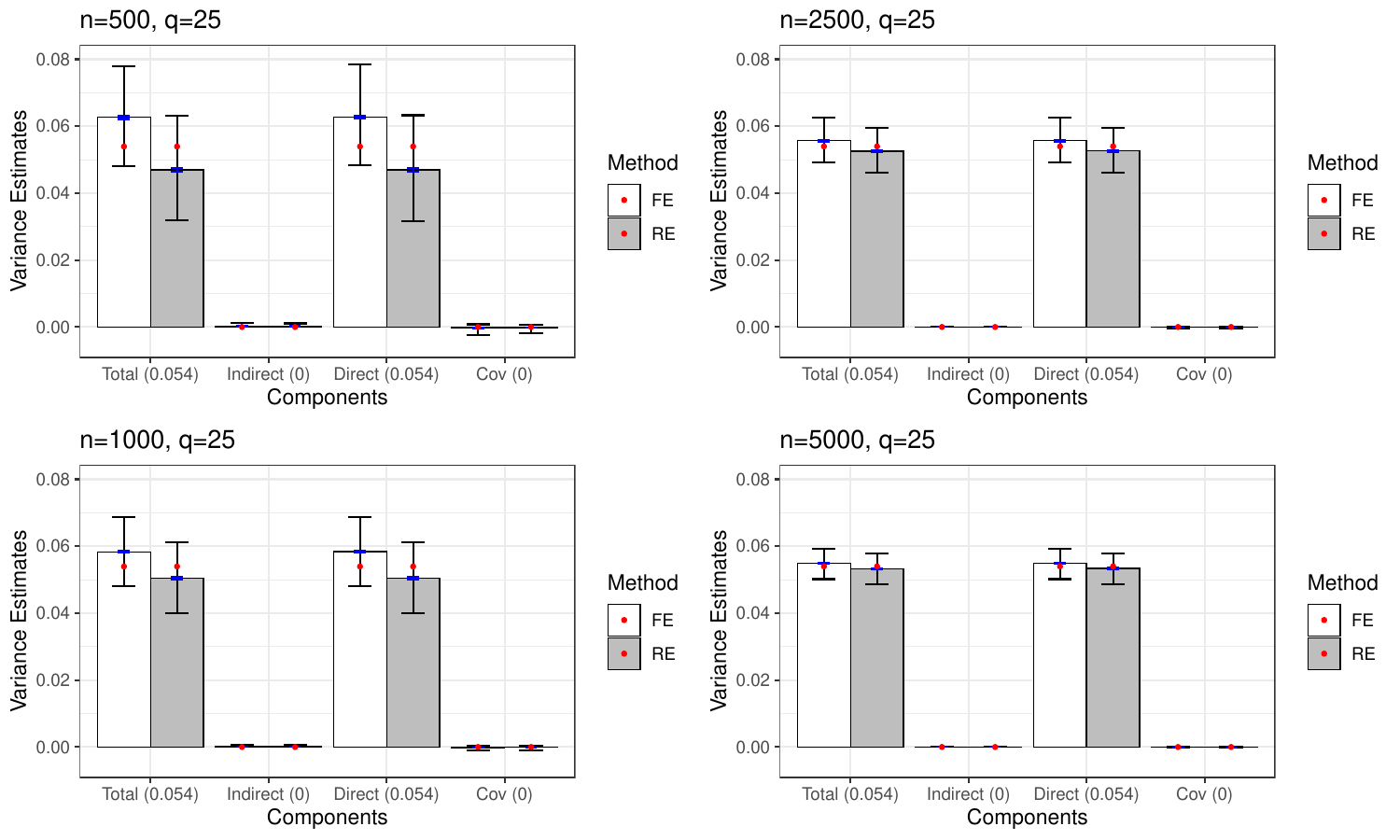}
\caption{Simulated sampling distribution means for the variance components under the fixed- and random-effect models for the binary outcomes under the different total number of patients $n$, based on 1000 replications. The direct mediator effect on the outcome was absent in the data generation. The red dots indicate the true variances. The $95\%$ quantile interval of the sampling distribution is represented by the black error bar. The $95\%$ confidence interval for the mean is represented by the blue error bar, reflecting the Monte Carlo error in the estimated mean of the sampling distribution.}
\label{Plot8}
\end{figure}

Figure \ref{Plot5} shows the simulated sampling distribution means for the total between-hospital variance, the indirect effect term, the direct effect term, and the covariance term for the binary outcomes, under different combinations of $n$ (total number of patients) and $q$ (total number of hospitals), and with $\sigma=0$, based on 1000 replications. From the plots, we can observe that the performance of the estimators is good for the scenarios where the per hospital number of patients is sufficiently large. However, some small sample bias in the estimated direct effect presents (overestimated in the fixed-effect model and underestimated in the random effect model) with the small number of patients per hospital (top-right panel 500/25=20 patients per hospital on average), which disappears with the increasing number of patients. This is more pronounced for binary compared to continuous outcomes. The density plots in figure \ref{Plot6} show that the uncertainty of the indirect effect and direct effect terms is driven by the total number of patients, with convergence towards the true value observable when $n$ increases.  From figure \ref{Plot7}, we observe that the covariance term estimates are close to 0 when $\sigma=0$ in the data generating mechanism, non-zero when $\sigma=2$ in the data generating mechanism. From figure \ref{Plot8}, we again observe that the estimated direct effect term and the estimated covariance term disappear when the direct mediator effect on the outcome was absent ($\beta_{q+1}=0$) in the data generation.

\section{Illustration in real data}\label{section:illustration}

To demonstrate and test the methods, we used population-level Ontario datasets housed at the Institute for Clinical Evaluative Sciences (ICES) on kidney cancer care, which comprises more than 10,000 nephrectomy patients in Ontario identified from the hospital Discharge Abstract Database (DAD) from 1995-2014. This dataset was cross-linked to kidney cancer diagnosis in the Ontario Cancer Registry and corresponding Ontario Health Insurance Plan (OHIP) billings, as well as to abstracted pathology reports. There is evidence to show that the early-stage radical nephrectomy patients with minimally invasive surgery experienced a shorter length of stay than the patients with open surgery \citep{Alice:2015, Luciano:2016, Daignault:2019}. Our aim was to determine how much of the between-hospital variation in the length of stay is due to the proportion of the minimally invasive surgery (MIS) performed and how much is due to all other hospital-level practices.

We identified a cohort of 4139 early-stage (T1-T2) radical nephrectomy patients treated in 72 different hospitals in Ontario who had complete data on the covariates (age, sex, income quintile, Charlson comorbidity score, ACG comorbidity score, presence of chronic kidney disease risk factors, days from diagnosis to treatment, year of diagnosis, tumor size, T stage). The average of the log value of the length of stay was 1.67 with a range from 0 to 5.12, while the overall proportion of patients receiving minimally invasive surgery was $63.82\%$. For the hospital assignment model, we used R package VGAM \citep{vgam} to fit a multinomial logistic assignment model for the hospital assignment. Because this model is only used for weighting the variance contributions rather than for case-mix adjustment, we estimated only intercept terms for hospitals that had treated less than 40 patients over the study period to avoid empty covariate categiries for the small volume hospitals. For the log-transformed outcome, we used R package lme4 \citep{lme4} to fit a mixed-effect linear model, adjusting for the mediator MIS and aforementioned covariates. For the binary mediator model, we fitted a mixed-effect logistic regression model adjusting the aforementioned covariates.

Table \ref{table_case_mix} shows a summary table of the effects of case-mix factors for the log value of Length of stay (outcome model) and MIS (mediator model). Here $\chi^2$ represents the likelihood ratio test for the between-hospital effects (random effect variance with one degree of freedom). The $\chi^2$ values in the likelihood ratio test indicate statistically significant between-hospital variation in Length of stay and MIS. The numbers in the table are the regression coefficients (t-values) for the log value of Length of Stay and log-odds ratios (z-scores) for the MIS. MIS has very strong negative effects on the length of stay, which provides evidence that the early-stage radical nephrectomy patients with minimally invasive surgery experienced a shorter length of stay than the patients with open surgery. Also, older patients with higher Charlson scores had a longer average length of stay after radical nephrectomy. Moreover, patients diagnosed later had a higher chance to receive minimally invasive surgery, and experienced a shorter average time stay at the hospital after the surgery. Furthermore, patients with a larger tumor size had a higher chance of receiving open surgery rather than minimally invasive surgery.

\begin{table}[!ht]
\centering
\begin{adjustbox}{max width=130mm}
\begin{tabular}{lcc}
\hline
                      & log Length of stay (outcome model) & MIS (mediator model) \\
Covariate             & $n=4139$                       & $n=4139$             \\
                      & $q=72$                         & $q=72$             \\
                      & $\chi^2=29$                    & $\chi^2=1078$        \\ \hline
MIS                   & $-0.27 (-18.40)$               &                      \\
Male sex              & $-0.02 (-1.33)$                & $0.07 (0.89)$        \\
Age                   & $0.01 (11.42)$                 & $-0.01 (-2.03)$      \\
Income quintile       & $-0.02 (-3.67)$                & $0.01 (0.26)$        \\
Rural vs urban        & $-0.01 (-0.47)$                & $-0.14 (-1.13)$      \\
Charlson score        & $0.04 (10.00)$                 & $-0.11 (-4.38)$      \\
ACG score             & $<|0.01| (3.02)$               & $0.01 (1.81)$        \\
log (dx to tx days+1) & $0.02 (4.28)$                  & $0.02 (1.02)$        \\
Year of dx            & $-0.02 (-8.77)$                & $0.20 (14.94)$       \\
Tumor size (cm)       & $0.01 (2.24)$                  & $-0.16 (-7.12)$      \\
CKD risk factors      & $0.01 (0.99)$                  & $-0.03 (-0.32)$      \\
T2 vs T1 stage        & $0.07 (3.17)$                & $-0.35 (-2.45)$        \\ \hline
\end{tabular}
\end{adjustbox}
\caption{A summary table of the effects of case-mix factors for the log value of Length of stay (outcome model) and MIS (mediator model). The numbers are the regression coefficients (t-values) for the log value of Length of Stay and the log-odds ratios (z-scores) for the MIS, while $n$ and $q$ represent the number of patients and the number of hospitals. $\chi^2$ represents the likelihood ratio test for the between-hospital effects (one degree of freedom). dx stands for diagnosis and tx for treatment (surgery).}
\label{table_case_mix}
\end{table}

Table \ref{mediation_realdata} shows how the estimated between-hospital variance in Length of stay decomposes into mediated through the MIS (indirect effect), direct effect through the all other pathways, and the covariance term between indirect and direct effect, along with the corresponding $95\%$ credible interval using the approximate Bayesian method described in Section \ref{Variance_estimation}. From the results, we observe that most of the between-hospital variation is due to mediation through MIS ($50.86\%$). The direct effect ($36.21\%$) and the covariance between direct and indirect effects ($12.93\%$) had relatively smaller contribution. These results are consistent with the hospital-specific results of \citet{Daignault:2019}, but with the proposed approach we are able to summarize the hospital-specific mediation with a single mediation decomposition for the between-hospital variance.

\begin{table}[!ht]
\centering
\begin{adjustbox}{max width=120mm}
\begin{tabular}{cccccc}
                   &           & \multicolumn{4}{c}{Source of variation}                                   \\ \hline
Process/outcome    &           & Total            & Direct effect    & Indirect effect  & Covariance       \\ \hline
MIS/Length of Stay & Variance  & $0.0116$ $(100.00\%)$          & $0.0042$ $(36.21\%)$          & $0.0059$ $(50.86\%)$          & $0.0015$ $(12.93\%)$          \\
                   & $95\%$ CI & $(0.0092, 0.0150)$ & $(0.0033, 0.0053)$ & $(0.0040, 0.0084)$ & $(<0.0001, 0.0029)$ \\ \hline
\end{tabular}
\end{adjustbox}
\caption{The estimated values of the between-hospital variance in Lenght of stay decomposed into mediated through the MIS (indirect effect), remaining variation due to all other pathways (direct effect), and the covariance between indirect and direct effect. Note: 95\% CI= 95\% credible interval.}
\label{mediation_realdata}
\end{table}

\section{Discussion}\label{section:discussion}

To determine whether the performance differences between hospitals are explained by differences in a given process of care, we formulated a causal mediation analysis decomposition for total between-hospital variance. Although the decomposition is not a true variance decomposition (as it involves a potentially negative term), each component in this decomposition is interpretable. Such a mediation analysis enables targeting quality of care interventions to processes that show mediation. Compared with previously proposed effect size or $R^2$ type measures for mediated effects in the structural linear modeling framework, our decomposition is causally interpretable as it is defined using the potential outcomes notation, and the estimation methods allow the choice of link function and can accommodate exposure-mediator interactions. 

For the estimation of the causal variance mediation decomposition, we used both random and fixed effect models. Because both of these were correctly specified in our simulation study, the two modeling approaches performed similarly in large samples, demonstrating consistency of the estimators. However, some differences were observed in small samples, especially when the number of patients per hospital was small (in the $n=500$ and $q=25$ scenario this was only 20). In such scenarios with the binary outcome, the estimates for the direct effect component indicated some bias, which was smaller with the random effect model. The uncertainty in the estimates was also smaller with the random effect model. This suggests that the added shrinkage has a stabilizing effect on the hospital effect estimates when small volume hospitals are present and there is limited information to estimate the hospital-specific coefficients in a fixed effects model. On the other hand, the random effect model may not produce consistent estimates if the random effect distribution is misspecified. However, we did not test misspecification scenarios herein. In the mediation analysis the choice between fixed and random effects can also be different for the outcome and mediator models, i.e. it is possible to specify a random effect model for one of these and a fixed effects model for the other. As noted in Section \ref{section:generation}, we used random effect models for estimation purposes, while true causal quantities were defined in terms of fixed effects. Thus, we did not explicitly try to establish a connection between the causal variance components and random effect variance parameters, although we noted a certain connection in a particular special case in Section \ref{section:true}. Generalizing this connection could suggest measures that are based on transformations of the estimated random effect variances, analogous to the median odds ratio \citep{larsen2005appropriate}. We leave this to be pursued in further work.

Here we focused on mediation analysis with a dichotomous mediator. However, in principle the approach generalizes to continuous mediators, as in Equation \eqref{equation:mediator}. Instead of specifying parametric models for the mediator, in this case we might be interested in more flexible distribution free model specifications. For this purpose, the semi-parametric model of \cite{Liu:2017} based on ordinal regression might provide an alternative. Because of the hospital comparison context, herein we focused on multi-category categorical exposures. However, in principle the variance decomposition approach is also generalizable to continuous exposures.

There are several possible extensions based on the current research. Effect decompositions can be formulated for multiple causal mediators that are either sequential or causally unrelated \citep{Lange:2014,Loeys:2017}. Methods have also been proposed in the literature for the analysis of high-dimensional mediators, usually based on latent variable dimension reductions as mediators, where the resulting the multiple mediators are unrelated to each other \citep{Andriy:2019}. In the hospital comparison context we might be interested in multiple sequential mediators representing different processes on the care pathway.

As noted in Section \ref{section:intro}, the present problem is distinct from multilevel mediation analysis, since herein the clusters (hospitals) are directly used as levels of the categorical exposure. However, in the hospital comparisons context further clustering may be present for example through surgeons nested within the hospitals. In \citet{chen2020hierarchical} we proposed a four-way hierarchical causal variance decomposition for settings where such a nested exposure hierarchy is present. Because the within-hospital between-surgeon variance component appears in this decomposition as a separate component of the between-hospital variance (i.e. it is entirely part of the residual variance of the three-way decomposition discussed in the present work), it is possible to further decompose the between-surgeon variance component in the mediation analysis sense using the same approach we proposed herein. Separate mediation analyses would then be carried out at the hospital and at the surgeon level.

\section*{Acknowledgements}
This work was supported by a Discovery Grant from the Natural Sciences and Engineering Research Council of Canada (to OS), a Catalyst Grant in Health Services and Economics Research from the Canadian Institutes of Health Research (to AF, KAL and OS) and the Ontario Institute for Cancer Research through funding provided by the Government of Ontario (to BC).

This study contracted ICES Data \& Analytic Services (DAS) and used de-identified data from the ICES Data Repository, which is managed by ICES with support from its funders and partners: Canada’s Strategy for Patient-Oriented Research (SPOR), the Ontario SPOR Support Unit, the Canadian Institutes of Health Research and the Government of Ontario. The opinions, results and conclusions reported are those of the authors. No endorsement by ICES or any of its funders or partners is intended or should be inferred.

Parts of this material are based on data and information compiled and provided by CIHI. However, the analyses, conclusions, opinions and statements expressed herein are those of the author, and not necessarily those of CIHI.

Parts of this material are based on data and information provided by Cancer Care Ontario (CCO). The opinions, results, view, and conclusions reported in this paper are those of the authors and do not necessarily reflect those of CCO. No endorsement by CCO is intended or should be inferred.

\end{document}